\documentclass[12pt,journal,draftcls,letterpaper,onecolumn]{IEEEtran}
\usepackage{amssymb}
\usepackage{amsmath}
\usepackage{epsfig}
\usepackage{algorithm}
\usepackage{algorithmic}
\usepackage{graphicx}
\usepackage{subfigure}
\usepackage{bm}
\usepackage{amsthm}
\usepackage{multirow}
\usepackage{tikz}
\usetikzlibrary{arrows.meta}
\usetikzlibrary{bending}
\usetikzlibrary{topaths}

\allowdisplaybreaks

\begin{document}

\title{A Unified Pilot Design for Integrated Sensing and Communications}
\author{\IEEEauthorblockN{Pu~Yuan\\}
\IEEEauthorblockA{Eagle Labs, TCL New Technologies, China 201203\\}
Email: pu.yuan@tcl.com}
\maketitle
\begin{abstract}
This paper investigates a unified pilot signal design in an orthogonal frequency division modulation (OFDM)-based integrated sensing and communications (ISAC) system. The novel designed two-dimensional (2D) pilot signal is generated on the delay-Doppler (DD) plane for sensing, while its time-frequency (TF) plane transformation acts as the demodulation reference signal (DMRS) for the OFDM data. The well-designed pilot signal preserves orthogonality with the data in terms of resource occupancy in the TF plane and quasi-orthogonality in terms of codeword in the DD plane. Leveraging these nice properties, we are allowed to implement sensing detection in the DD plane using a simple 2D correlation, taking advantage of the favorable auto-correlation properties of the 2D pilot. In the communication part, the transformed pilot in the TF plane serves as a known DMRS for channel estimation and equalization. The 2D pilot design demonstrates good scalability and can adapt to different delay and Doppler resolution requirements without violating the OFDM data detection and can overcome the fractional Doppler with limited sensing resources. Experimental results show the effective sensing performance of the proposed pilot, with only a small fraction of power shared from the OFDM data, while maintaining satisfactory symbol detection performance in communication.
\end{abstract}


\section{Introduction}\label{sec:sec0}
In recent years, we have witnessed a proliferation of research works in ISAC systems. The concurrent ISAC systems can be categorized into two main types: data-based sensing and pilot-based sensing. The latter exhibits better resilience to interferences in multi-user systems, while being designed to adapt multiple-station sensing.

Most of the literature considers the OFDM-ISAC system which can smoothly evolve from the 5G system \cite{7855671}, and it is proven to give acceptable sensing performance in various scenarios \cite{5599316}. However, as pointed in \cite{9557830}, the OFDM-ISAC system is prone to high Doppler such that a notable performance degradation may occurs in vehicular communications. The orthogonal time frequency space (OTFS) modulates the data symbols in the DD plane, shows a promising effort in combating with the time-varying channel especially in high mobility scenarios \cite{Hadani2017otfs}. As the OTFS-ISAC system is firstly proposed in \cite{8835764}, extensive works have discussed on the advantages of sensing in the DD domain \cite{9109735}. Although OTFS is recognized as a promising techniques in both sensing and communications, we may still face the fact that the OFDM may still be preferred in most scenarios for its simplicity and effectiveness.

The conventional DD plane pilot design, such as the commonly discussed impulse pilot \cite{Raviteja2019embedded}, exhibits a high peak-to-average power ratio (PAPR) problem. To overcome the PAPR problem, sequence based pilot that spreads the pilot power throughout the delay-Doppler plane design are provided in \cite{yuan2022lpp}, which efficiently reduce the magnitude variation of the time domain samples. In \cite{yuan2023us}, an ISAC system in which the the OFDM data are code division multiplexed (CDM) with a 2D pilot preserves good performance in both sensing and communications. 

In this paper, we propose a unified pilot design for ISAC systems in which the pilot is modulated in the DD plane to play its advantages in sensing detection, while the data part is still modulated in the TF plane for better compatibility and simplicity. The design philosophy of the 2D pilot signal follows the fact that the channel coupling effect on the transmitted signal occurs in the DD plane through a twisted convolution mechanism \cite{mohammed2022otfs}, where the received pilot signal exhibits dual-dimensional cyclic shifts and the phase changes. The major advantages of the proposed design include:
\begin{itemize}
\item[-] \textbf{\textit{The scalability.}} 
The dimension of the 2D pilot can be flexibly adjusted to fulfill the sensing demands without violating the numerology of the communication system, and the sensing detection complexity scales linearly with the dimension.

\item[-] \textbf{\textit{The quasi-orthogonality.}}
The 2D pilot is designed to keep low cross-correlation with the data, which guarantees a high signal-to-interference-plus-noise ratio (SINR) in sensing. 

\item[-] \textbf{\textit{The compatibility.}}
The comb-like pattern of the transformed pilot in TF plane is friendly to the legacy OFDM user for frequency domain channel estimation and equalization.  

\item[-] \textbf{\textit{The flexibility.} }
The 2D pilot corresponded to multi-antennas can be mapped in multiple layered to support angular estimation, and can be sparsely mapped for the purpose of energy saving.
\end{itemize}

\section{ISAC Signal Transmitter Design}\label{sec:sec1}

To leverage the twisted convolution in sensing, we design a sensing pilot with good auto and cross correlations, as well as a simple implanting scheme to achieve orthogonal multiplexing of the transformed pilot and OFDM data in the TF plane.

\subsection{Sensing Pilot Design and Property}\label{subsec:subsec0}

We firstly construct a 2D sensing pilot in the DD plane using two sequences $\mathbf{a}=[a_1,a_2,...,a_Q]$ and $\mathbf{b}=[b_1,b_2,...,b_P]^H$ who are with good auto/cross-correlation properties, i.e., $\forall 0\leq p,i<P, 0\leq q,j<Q, p,i,q,j\in \mathbb{N}$, i.e., 
$
\textrm{tr}\left(\frac{1}{Q}\mathbf{a}^H_{[q]}\mathbf{a}_{[j]}\right)
 = \frac{1}{Q}\mathbf{a}_{[q]}\mathbf{a}_{[j]}^H = 
\left\{
\begin{array}{cc}
1, & q=j,  \\
\epsilon_a & q\neq j.   \\
\end{array}\right.
$
and 
$
\textrm{tr}\left(\frac{1}{P}\mathbf{b}_{[p]}\mathbf{b}^H_{[i]}\right)
 = \frac{1}{P}\mathbf{b}^H_{[p]}\mathbf{b}_{[i]} = 
\left\{
\begin{array}{cc}
1, & p=i,  \\
\epsilon_b & p\neq i.  \\
\end{array}\right.
$
,
where $(\cdot)^H$ denotes the conjugate transpose, $(\cdot)_{[i]}$ denotes the cyclic-shift of a vector and $0\leq \left|\epsilon_a\right|, \left|\epsilon_b\right| \ll 1$. The 2D pilot can be generated from the Kronecker product or product of the $\mathbf{a}$ and $\mathbf{b}$ with arbitrary cyclic shift,
$
\mathbf{P}_{[q,p]} = \mathbf{a}_{[q]}\otimes \mathbf{b}_{[p]} = \mathbf{b}_{[p]}\mathbf{a}_{[q]},
$
where $(\cdot)_{[i,j]}$ denotes cyclic shifting $i$ and $j$ units in row and column. Based on above notations, the matrix correlation between $\mathbf{P}_{[j,i]}$ and $\mathbf{P}_{[q,p]}$ has the following property,
\begin{equation}
\frac{1}{QP}\langle \mathbf{P}_{[q,p]},\mathbf{P}_{[j,i]}\rangle = 
\left\{
\begin{array}{cc}
1, & q=j, p=i  \\
\epsilon_p & q=j, p\neq i   \\
\epsilon_q & q\neq j, p=i   \\
\epsilon_q\epsilon_p & q\neq j, p\neq i   \\
\end{array}\right.
\label{eq:eq3}
\end{equation}

\begin{figure}
\centering
\includegraphics[width=0.7\textwidth]{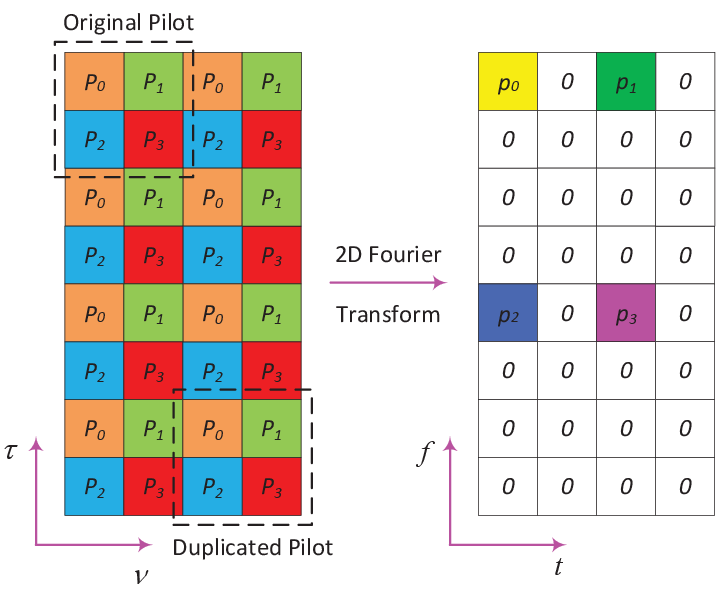}
  \caption[]{DD plane repetition results in TF plane zero interpolation. The comb-like mapping of transformed pilot has a similar structure as the conventional DMRS. }
	\label{fig:fig0}
\end{figure}

Upon obtaining the 2D pilot, we utilize the property of the Fourier transform to develop our TF plane reference signal implanting scheme similar as the idea in \cite{yuan2023hfs}. Suppose a time domain sequence $\mathbf{s}$ and a frequency domain sequence $\mathbf{x}$ is a pair of Fourier transform, then the time domain interpolation with $i-1$ zeroes is equivalent to the frequency domain repetition for $i$ times, i.e., 
$
\mathrm{IDFT} ([s_0, \underbrace{0, ...}_{i-1}, s_1, 0, ..., s_{l-1}, 0, ...])
 = [\underbrace{x_0, ..., x_{l-1}, x_0, ..., x_{l-1}, ..., x_0, ..., x_{l-1}}_{\mathrm{Duplicated\ for\ i\ times.}}].
$
By repeating the sensing pilot in either dimension for $i$ times, we let its transformation occupy $\frac{1}{i}$ of resources in TF plane, and the unoccupied which labeled as zero is used for data, as illustrated in Fig. \ref{fig:fig0}.

\subsection{ISAC Signal Generation}\label{sec:sec2}

We consider a frame size of $M\times N$, where $M$ and $N$ corresponds to the number of delay and Doppler taps in DD plane, as well as the number of OFDM symbols and subcarriers in the time frequency plane, respectively. The Doppler and delay are quantized as $\tau=\frac{T}{M}$ and $\nu=\frac{1}{N\Delta f}$. We denote the symbol time and subcarrier spacing of the OFDM as $T$ and $\Delta f$. We denote the discrete samples of the 2D pilot in DD plane as $\mathbf{P}\in\mathbb{C}^{M \times N}$, which is the cyclical extension of the based pilot block $\mathbf{P}_{dd}\in\mathbb{C}^{\frac{M}{d_{f}} \times \frac{N}{d_{t}}}$, where $d_t$ and $d_f$ are integers correspond to the DMRS densities in time and frequency. Then we apply the inverse symplectic fast Fourier transform (ISFFT) to obtain the transformed 2D pilot $\mathbf{P}_{tf}$ in TF plane as follows, which is given by,
\begin{equation}
\mathbf{P}_{tf} = \mathbf{F}_M \mathbf{P} \mathbf{F}^H_N = \mathbf{F}_M(\mathbf{1}_{1\times d_{t}}\otimes \mathbf{P}_{dd} \otimes \mathbf{1}_{d_{f} \times 1})\mathbf{F}^H_N.
\label{eq:eq5}
\end{equation}
, where $\mathbf{F}_M$, $\mathbf{F}^H_M$ denotes the M-point DFT/IDFT. 

\begin{figure}
\centering
\includegraphics[width=0.7\textwidth]{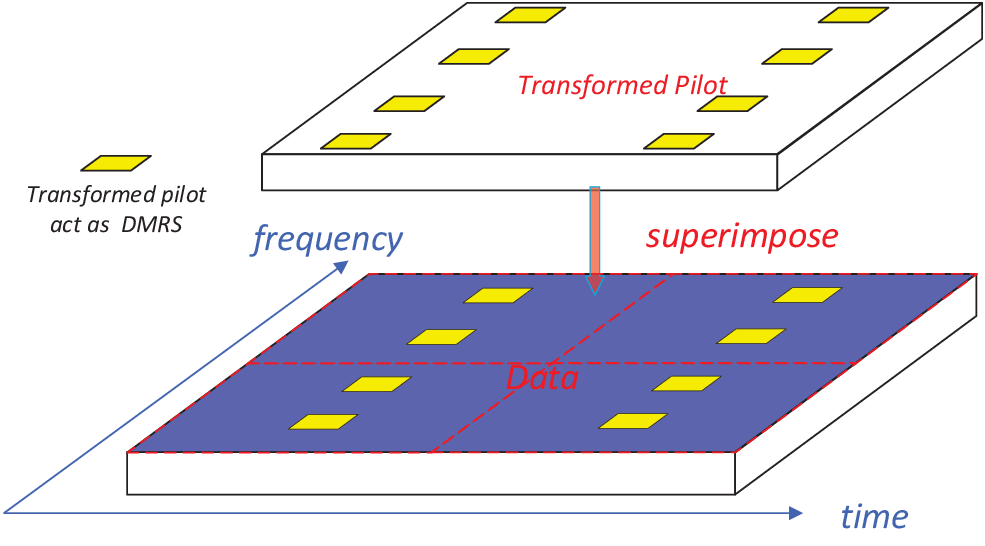}
  \caption[]{The transformed pilot in TF plane is naturally the DMRS, where OFDM data is mapped to the positions labeled with zero in figure \ref{fig:fig0}.}
	\label{fig:fig1}
\end{figure}

We assume the discrete samples of TF plane data as $\mathbf{D}\in\mathbb{C}^{M \times N}$, where some elements are punctured for DMRS placement. The transformed 2D pilot is implanted into the OFDM data symbols in the TF plane, i.e., 
$
\mathbf{X} = \mathbf{F}_M \mathbf{P} \mathbf{F}^H_N + \mathbf{D}.
$
as illustrated in figure \ref{fig:fig1}. 

The implanted TF plane samples are converted to the time-delay plane and appended with symbol-wise cyclic prefix (CP) as follows,
\begin{equation}
\tilde{\mathbf{X}} = \mathbf{B}_{cp}\mathbf{F}^H_M\mathbf{X} = \mathbf{B}_{cp}\left(\mathbf{P}\mathbf{F}^H_N + \mathbf{F}^H_M\mathbf{D}\right),
\label{eq:eq6}
\end{equation}
where $\mathbf{B}_{cp} = \left[
\begin{array}{ccc}
\mathbf{0}_{L_{cp}\times (M-L_{cp})} & \mathbf{I}_{L_{cp}}  \\
  \mathbf{I}_{M} &   \\
\end{array}
\right]$
, which is the operator for appending a CP of $L_{cp}$ samples \cite{9303350}. 
Then the discrete time signal $\tilde{\mathbf{x}}$ is give by vectorizing the time-delay plane samples and imposing a prototype filter $\mathbf{G}_{tx}$ , 
\begin{equation}
\tilde{\mathbf{x}} = \left(\mathbf{F}^H_N\otimes\mathbf{G}_{tx}\right)\mathbf{vec}\left(\mathbf{B}_{cp}\mathbf{P}\right) + \mathbf{vec}\left(\mathbf{G}_{tx}\mathbf{B}_{cp}\mathbf{F}^H_M\mathbf{D}\right),
\label{eq:eq7}
\end{equation}
where $\mathbf{vec}(\dot)$ denotes the vectorization operator, $G_{tx}$ and $G_{rx}$ denote the receive and transmit filters, respectively.

With rectangular pulse shaping we have $\mathbf{G}_{tx}=\mathbf{I}_M$. Without loss of generality, we limit our analysis with the discrete time model and abbreviate the discussion on continuous waveform and assuming perfect synchronization and best sampling.

\section{Sensing and Communication Receiver Processing}\label{sec:sec3}
At the sensing receiver, after ADC and down sampling, the DD plane received pilot $\mathbf{R}$ is actually an power-attenuated and cyclic-shifted version of $\mathbf{P}$, with additional IDI (inter-Doppler interferences) brought by the fractional Doppler. Denote $r[k,l]$ and $p[k,l]$ as the elements of $\mathbf{R}$ and $\mathbf{P}$ respectively, we can write the DD plane input-output relationship following the same line as \cite{Raviteja2018interference},
\begin{align}
\label{eq:eq8}
r[l,k] & \approx \sum^{L-1}_{i=0}\sum^{N_i}_{q=-N_i}h_i\alpha_{i}(l,k,q)(p[[l-l_i]_M \nonumber \\
& ,[k-k_i+q]_N])e^{j2\pi\frac{(L_{cp}+l-l_i)(k_i+\kappa_i)}{N(M+L_{cp})}}, 
\end{align}
where $h_i$ and $\kappa_i$ is the channel attenuation factor and fractional Doppler of the path $i$, and 
$\alpha_{i}(l,k,q)  = \left\{
\begin{array}{cc}
\frac{1}{N}\beta_i{q} & l_i\leq l<M  \\
\frac{1}{N}(\beta_i(q)-1)e^{-j2\pi\frac{[k-k_i+q]_N}{N}} &  0\leq l<l_i
\end{array}\right.
$
,
$
\beta_i(q) = \frac{e^{j\frac{2\pi}(-q-\kappa_i)}-1}{e^{j\frac{2\pi}{N}(-q-\kappa_i)}-1}
$
contribute to the interfering term due to the rectangular pulse shaping and fractional Doppler, where the impact of CP on the phase term is derived from the results in \cite{9303350}.

\subsection{Sensing Pilot Detection for Integer Delay and Doppler}\label{subsec:subsect1}

Note that although fractional Doppler is ubiquitous in practical wireless environments, in certain scenario it is reasonable to assume large $M$ and $N$ such that the resolution of delay and Doppler are fairly enough to approximate them to integers. Furthermore, we show later that a coarse sensing determining the integer delay and Doppler helps us to further analyze the fractional Doppler to refine the sensing results.  

If we neglect the impact of fractional Doppler, i.e., let $q=0$ and $\kappa_i=0$, (\ref{eq:eq8}) can be simplified as, 
\begin{equation}
r[k,l] = \sum^{L-1}_{i=0}{{h_i\alpha_{i}(k,l)(p[[l-l_i]_M,[k-k_i]_N])e^{j2\pi\frac{(L_{cp}+l-l_i)k_i}{N(M+L_{cp})}}}}, 
\label{eq:eq9}
\end{equation}
where 
$\alpha_{i}(l,k)  = \left\{
\begin{array}{cc}
1 & l_i\leq l<M  \\
\frac{N-1}{N}e^{-j2\pi\frac{[k-k_i]_N}{N}} &  0\leq l<l_i
\end{array}\right.
$
for $\beta_i(0)=N$. 
The approximated input-output model considering the integer delay and Doppler only helps us to obtain a coarse sensing result. Suppose in 
(\ref{eq:eq9}) each echo path corresponds to one sensing object, then the sensing channel corresponds to this object is,
\begin{equation}
r_i[l,k] = h_i\alpha_{i}(l,k)(p[l,k]\ast \delta[[l-l_i]_M,[k-k_i]_N])e^{j2\pi\frac{(L_{cp}+l-l_i)k_i}{N(M+L_cp)}}, 
\label{eq:eq10}
\end{equation}
where $\alpha_{i}(l,k)$ is the time-varying phase change caused by the Doppler and $p[l,k]\ast \delta[[l-l_i]_M,[k-k_i]_N]$ is the cyclic-shift operation to $\mathbf{P}$.

By ignoring the attenuation factor $h_i$ which is irrelevant to the integer delay and Doppler, we approximately give the equivalent matrix form input-output of the echo path $k_i,l_i$ as,
\begin{equation}
\mathbf{R} = \mathbf{P}_{[l_i,k_i]}\odot \Xi_{l_i,k_i}, 0\leq l_i<\frac{M}{d_f}, 0\leq k_i<\frac{N}{d_t}, l_i,k_i\in \mathbb{N},
\label{eq:eq11}
\end{equation}
where $\odot$ denotes the Hadamard product, i.e., the point-wise product. In (\ref{eq:eq11}), $\Xi_{l_i,k_i}$ is the phase offset matrix applied to DD plane discrete samples with the $\alpha_{i}(l,k)e^{j2\pi\frac{(L_{cp}+l-l_i)k_i}{N(M+L_{cp})}}$ as its entries. 
The sensing detector is given by finding the maximum inner product between $\mathbf{R}$ and a locally matching pilot $\mathbf{P}_{[l_j,k_j]}$, 
\begin{align}
& \mathop{\arg\max}_{l_j,k_j} \langle \mathbf{P}_{[l_i,k_i]},\mathbf{P}_{[l_j,k_j]}\rangle \nonumber \\ 
& = \sum^N_{n=1}{\sum^M_{m=1}{\left[\mathbf{P}_{[l_i,k_i]}\odot \mathbf{P}_{[l_j,k_j]}\right]{(n,m)}}} |_{l_i=l_j,k_i=k_i}\nonumber \\ 
& = \sum^N_{n=1}{\sum^M_{m=1}{\left[\mathbf{R}\odot \Xi^{*}_{l_i,k_i}\odot \mathbf{P}_{[l_j,k_j]}\right]{(n,m)}}}|_{l_i=l_j,k_i=k_i}.
\label{eq:eq12}
\end{align}
Note that equation (\ref{eq:eq12}) approaches the maximum when $l_i=l_j$ and $k_i=k_j$. It also explicitly gives the structure of the sensing detector, which can be summarized as a series of hypothesis tests of the phase-compensated received signal samples with the cyclic shifted version of the original 2D pilot in the DD plane. 

For sensing detection, we usually considers the line-of-sight (LOS) path denoted by $i=0$ to the sensing target, while the non-Line-of-Sight (NLOS) path or path to other reflectors as the interfering paths. Hence the phase-compensated received signal can be decomposed into three components,
\begin{align}
&\mathbf{R} \odot \Xi^{*}_{l_i,k_i}  = \sum^{L-1}_{i=0}{h_i(\mathbf{D}_{[l_i,k_i]}+(\mathbf{F}^H_M\mathbf{P}\mathbf{F}_N)_{[l_i,k_i]})} + \mathbf{W} \nonumber \\
&= \underbrace{h_0\mathbf{P}_{[l_0,k_0]}}_{Signal} 
+ \underbrace{\sum^{L-1}_{i=1}{h_i\mathbf{P}_{[l_i,k_i]}}+\sum^{L-1}_{i=0}{h_i(\mathbf{F}^H_M\mathbf{D}\mathbf{F}_N)_{[l_i,k_i]}}+ \mathbf{W}}_{Inter\quad Path\quad Interference\quad (IPI)\quad and\quad Noise}
\label{eq:eq13}
\end{align}

As described in section \ref{sec:sec2}, the sensing detection is implemented by feeding the received DD plane signal $\mathbf{R}$ and $\mathbf{P}_{[l,k]}$ into a two-dimensional correlator, and the correlation peak occurs when the locally generated matrix $\mathbf{P}_{[l_0,k_0]}$ is chosen as an input. In such case, the output is given by  
\begin{align}
 \langle \mathbf{P}_{[l_0,k_0]},\mathbf{Y}\odot \Xi^{*}_{l_0,k_0}\rangle \triangleq MNh_0+\sum^{L-1}_{i=1}h_i\langle \mathbf{P}_{[l_0,k_0]},
\nonumber \\
\mathbf{P}_{l_i,k_i}\rangle +\sum^{L-1}_{i=0}{h_i\langle \mathbf{P}_{[l_0,k_0]},\mathbf{D}_{l_i,k_i}}\rangle + \langle \mathbf{P}_{[l_0,k_0]},\mathbf{W}^H \rangle.
\label{eq:eq14}
\end{align}
It is noted that in the above equation the value of correlation peak $MNh_{0}$ is distorted by the interference and noise terms. 
Denote $\vartheta_{0i}$, $\rho_{0i}$ and $\varsigma_{0i}$ as 
$\langle \mathbf{P}_{[l_0,k_0]},\mathbf{P}_{l_i,k_i}\rangle$, 
$\langle \mathbf{P}_{[l_0,k_0]},\mathbf{D}_{l_i,k_i}\rangle$ and 
$\langle \mathbf{P}_{[l_0,k_0]},\mathbf{W}^H\rangle$ respectively, we define the sensing SINR here as 
$
Z \triangleq \frac{MNh_0}{\sum^{L-1}_{i=1}{h_i\vartheta_{0i}} +\sum^{L-1}_{i=0}{h_i\rho_{0i}}+\varsigma_{0i}}.
$
Here, $Z$ characterizes the distortion level. Note that $\vartheta_{0i}$ can be neglect due to (\ref{eq:eq3}), and $\rho_{0i}$ and $\varsigma_{0i}$ are also insignificant as shown in \cite{yuan2023us}. Then we can claim that $Z$ is lower bounded and asymptotically increases to infinity when $MN$ goes to infinity.

The proposed integer delay and Doppler detection is fairly simple as it requires only shift and product-sum operations. It can be further simplified by abbreviating the phase compensation if phase noise resilient sequences are utilized, which is different from the channel estimation where the phase information is required for the symbol detection. 

\subsection{Magnitude Based Fractional Doppler Detection}\label{subsec:subsect2}

\begin{figure}
\centering
\includegraphics[width=0.7\textwidth]{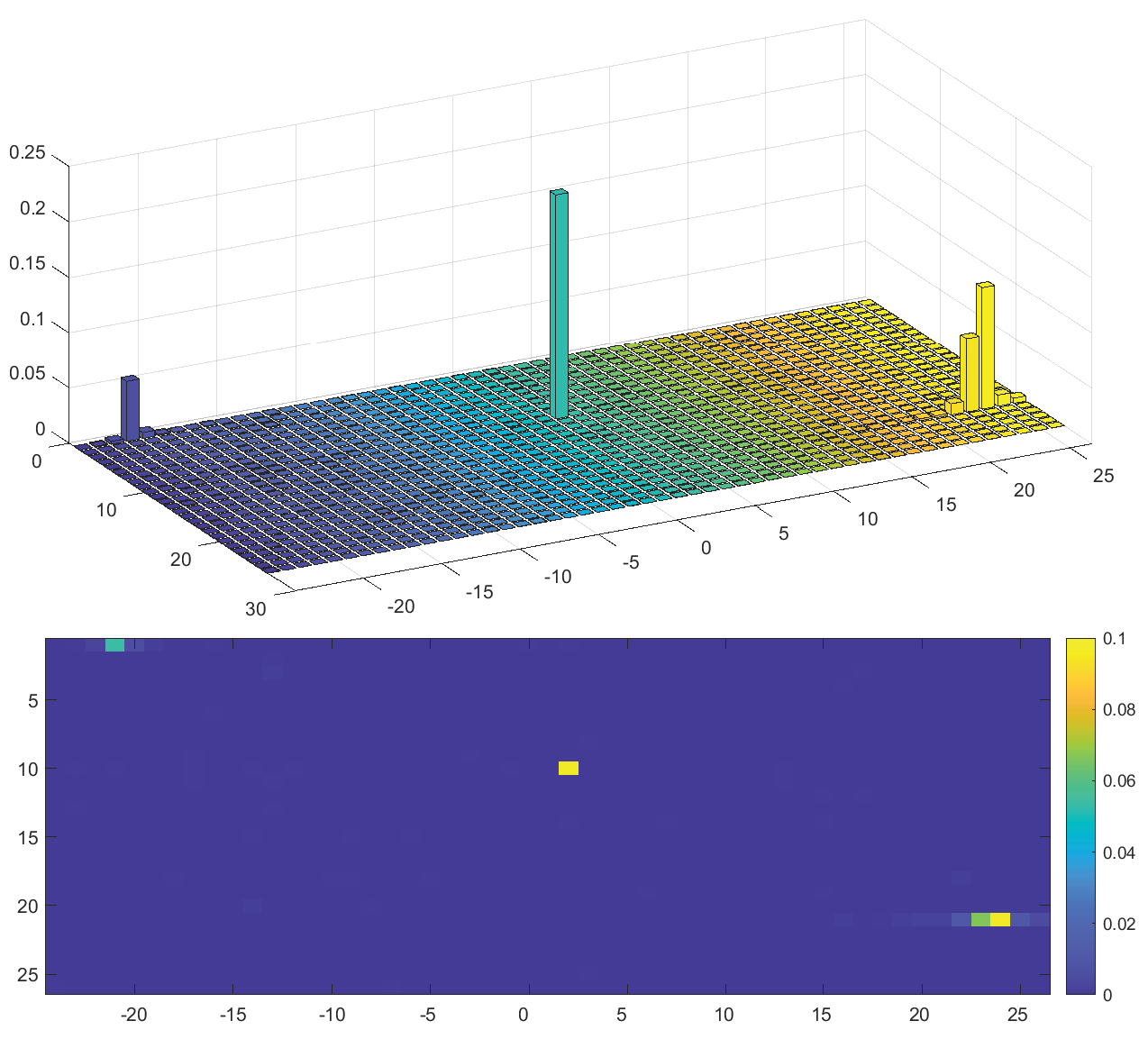}
  \caption[]{Illustration of a three targets detection. The one with larger fractional Doppler results in significant power dispersion to adjacent Doppler bins.}
	\label{fig:fig2}
\end{figure}

After initial detection all the delay paths are resolvable, which facilitates us to analyze the fractional Doppler path-by-path. 
Figure \ref{fig:fig2} illustrates the correlation output of three echo paths with $(\tau_i,\nu_i,h_i)$ as $(0,-21.7875,0.2352+0.3241i)$, $(9,0.9816,0.1336+0.7132i)$ and $(20,22.5537,0.5972+0.3935i)$. The multiple peaks depict the power dispersion due to the mismatches between the actual fractional Doppler and the assumed integer Doppler in the correlation.

This fact motivated us to explore the relationship between the magnitude information and the fractional Doppler value. Suppose the LOS path labeled with $i=0$ corresponds to the sensing target, then the echo signal $r_0$ is deduced from (\ref{eq:eq8}) as, 
\begin{align}
 & r_0[l,k] = \sum^{N_i}_{q=-N_i}h_0\alpha_{0}(l,k,q)(p[[l-l_0]_M], \nonumber \\
 & [k-k_0+q]_N) e^{j2\pi\frac{(L_{cp}+l-l_0)(k_0+\kappa_0)}{N(M+L_{cp})}} \nonumber \\
 & = h_0\alpha_{0}(k,l)p[[l-l_0]_M,[k-k_0]_N])e^{j2\pi\frac{(L_{cp}+l-l_0)(k_0)}{N(M+L_{cp})}} \nonumber \\
 & +\sum^{N_i}_{q=-N_i,q\neq 0}h_0\alpha_{0}(k,l,q)(p[[l-l_0]_M,[k-k_0+q]_N]) \nonumber \\
 & e^{j2\pi\frac{(L_{cp}+l-l_0)(k_0+\kappa_0)}{N(M+L_{cp})}}.
\label{eq:eq16}
\end{align}

Another observation from figure \ref{fig:fig2} is the power dispersion of the correlation peak is mainly caused by the fractional Doppler and the impact of IPI and noise term is trivial. Therefore we can neglect the IPI and noise term in (\ref{eq:eq13}), and substitute (\ref{eq:eq16}) in it to get the input-output in presence of fractional Doppler,
\begin{align}
& \langle \mathbf{P}_{[l_0,k_0]},\mathbf{Y}\odot \Xi^{*}_{[l_0,k_0]}\rangle \approx \sum_{k=1}^{N}\sum_{l=1}^{M}p[[l-l_0]_M,[k-k_0]_N] \nonumber \\
& r_0[l,k]\alpha^{*}_{0}(k,l)e^{-j2\pi\frac{(L_{cp}+l-l_0)k_0}{N(M+L_{cp})}} \nonumber \\
& = h_0 \sum_{k=1}^{N}\sum_{l=1}^{M}h_0 p[[l-l_0]_M,[k-k_0]_N] p[[l-l_0]_M, \nonumber \\
& [k-k_0]_N]
 + p[[l-l_0]_M,[k-k_0]_N]\alpha^{*}_{0}(l,k)\nonumber \\ 
& \sum^{N_i}_{q=-N_i,q\neq 0}\alpha_{0}(l,k,q)p[[l-l_0]_M,[k-k_0+q]_N] \nonumber \\
& e^{j2\pi\frac{(L_{cp}+l-l_0)\kappa_0}{N(M+L_{cp})}}
= MNh_0 + h_0e^{j2\pi\frac{\kappa_0}{M+L_{cp}}}\nonumber \\
& \sum_{k=1}^{N}\sum_{l=1}^{M}e^{j2\pi\frac{(L_{cp}+l-l_0)}{N}}\alpha^{*}_{0}(l,k)p[[l-l_0]_M,[k-k_0]_N] \nonumber \\
& \left(\sum_{q=-N_i,q\neq 0}^{N_i}{a_0(l,k,q)p[[l-l_0]_M,[k-k_0+q]_N]}\right).
\label{eq:eq17}
\end{align}

For simplicity we rewrite (\ref{eq:eq17}) as,
\begin{equation}
v_{l_0,k_0} = MNh_0 + \mu(l_0,k_0,\kappa_0)h_0.
\label{eq:eq17_plus}
\end{equation}
Note that the following variables are already known in (\ref{eq:eq17}), 
\begin{itemize}
\item[-] $p[k,l]$ is the element of the known 2D pilot $\mathbf{P}$. 
\item[-] $k_0$ and $l_0$ obtained as the integer delay and Doppler value.
\item[-] $M$, $N$ and $N_i$ are chosen as required.
\item[-] $v_{l_0,k_0}$ is calculated by the inner product of received signal and the local pilot.
\end{itemize}

Therefore, the only left unknown variables are the channel attenuation factor $h_0$ and fractional Doppler $\kappa_0$, respectively. Since $h_0$ is constant in each delay path, we can use the ratio of multiple adjacent peaks to cancel $h_0$. 

For example, by calculating the ratio of the most significant $N_j$ side peaks adjacent to $v_{l_0,k_0}$, i.e., $v_{l_0,k_0+N_j}$ and $v_{l_0,k_0-N_j}$, we construct a fractional Doppler detector on the delay and Doppler bin $(k_0,l_0)$ as,
\begin{align}
v_{det}(l_0,k_0,\kappa) = \frac{v_{l_0,k_0}}{\sum_{\Delta k=-N_j}^{N_j}v_{l_0,k_0+\Delta k}}
,
\label{eq:eq18}
\end{align}
where the only unknown variable $\kappa_0$ can be easily determined by an exhausted search for the one who gives the minimum Euclidean distance between the output of $v_{\mathrm{det}}$ and the measured results,
\begin{equation}
\kappa_0 = \arg \min_{\kappa \in (-0.5,0.5)} \left|v_{det}(l_0,k_0,\kappa) - v_{mes} \right|,
\label{eq:eq19}
\end{equation}
where $v_{\mathrm{mes}} = \frac{m(l_0,k_0)}{\sum_{k_i=k_0-1}^{k_0+1}m(l_i,k_i)}$ and $m(l_0,k_0)$ is the measured magnitude value at delay and Doppler bin $(l_0,k_0)$. Subsequently a two stage sensing detection algorithm is provided in \ref{alg:alg1}.

\begin{algorithm}
\textbf{Initialization} : maximum delay and Doppler taps $l_{max}$, $k_{max}$, $\mathbf{P}$, threshold $\gamma$, step size $\delta_{\kappa}$

\textbf{STEP-1: Integer Delay and Doppler Detection} 
\begin{itemize}
\item[a)] $\forall l_i\leq l_{max}, k_i\leq k_{max}$, construct the detection matrix $\mathbf{P}_{\mathrm{det}}$ for each integer delay and Doppler bin $(l_i, k_i)$ as 
$
\mathbf{P}_{\mathrm{det}} = \mathbf{P}_{[l_i,k_i]}\odot \Xi^*_{l_i,k_i}
$
\item[b)] Calculate the absolute value of matrix correlation, 
\begin{align}
v_{l_i,k_i} = \left|\sum^N_{n=1}{\sum^M_{m=1}{\left[\mathbf{R}\odot \mathbf{P}_{\mathrm{det}}\right]_{(n,m)}}}\right|. \nonumber
\label{eq:pd0}
\end{align}

\item[c)] 
$\forall l_j, k_j, v_{l_j,k_j} \geq \gamma$, a set of echo paths $\mathbf{p}=\{(l_j,k_j)\}$ are identified, i.e., a set of objects' relative velocity and distance can be determined.
\end{itemize}

\textbf{STEP-2: Fractional Doppler Estimation} :
\begin{itemize}
\item[a)] 
$\forall (l_j, k_j)\in \mathbf{p}$, construct the fractional Doppler detectors $v_{\mathrm{det}}(l_j,k_j,\kappa)$ according to (\ref{eq:eq17}), (\ref{eq:eq18}) and (\ref{eq:eq19}).

\item[b)] 
\begin{algorithmic}
		\WHILE {$n \leq \lfloor \frac{1}{\delta_{\kappa}} \rfloor$}
		\STATE $\eta(n) = \left|v_{\mathrm{det}}(l_j,k_j,-0.5+n\delta_{\kappa}) - v_{\mathrm{mes}} \right|$
		\STATE $\eta(n) = \max{(\eta(n),\eta(n-1))}$
		\STATE $\kappa_j = n\delta_{\kappa}$
		\STATE $n = n+1$
		\ENDWHILE
\end{algorithmic}

\item[c)] 
The final detected Doppler is $k_j+\kappa_j$, then the velocity is updated consequentially.
\end{itemize}

\label{alg:alg1}
\caption{Sensing Detection Based on 2D Pilot}
\end{algorithm}

\subsection{Communication Receiver}
The received signal after ADC and down sampling yields the discrete time samples, where the equivalent time domain input-output in (\ref{eq:eq8}) of each OFDM symbol can be written in vector form as $\mathbf{r} = \mathbf{H_t}\mathbf{s}+\mathbf{w}$ where $\mathbf{H_t} = \sum^L_{i=1}{h_i\bm{\Pi}^{l_i}\bm{\Delta}^{(k_i)}}, \mathbf{H_t}
 \in \mathbb{C}^{M\times M}$ is the equivalent time domain channel matrix derived from path delay, Doppler, channel gain and shaping pulse in the same way as \cite{Raviteja2018interference}. Hence the received samples per OFDM symbol after removing CP is given by,
\begin{equation}
\mathbf{r} = \mathbf{H_t}\left(\mathbf{p}\mathbf{F}^H_N+\mathbf{F}^H_M\mathbf{d}\right)+\mathbf{w},
\label{eq:eq20}
\end{equation}
where $\mathbf{p}$ and $\mathbf{d}$ are the columns of $\mathbf{P}$ and $\mathbf{D}$, respectively.

The transformed 2D pilot yields a comb-like mapping in the TF plane and is naturally the DMRS for channel estimation. This nice property maintains the backward compatibility with legacy OFDM users who adopt the symbol-wise frequency domain channel estimation and single-tap equalization to revert the modulation symbol. If the minimum mean squared error (MMSE) criterion is adopted, the OFDM symbol detector is given by,
\begin{equation}
\hat{\mathbf{x}} = (\mathbf{h}^{*}_{tf}\odot (\mathbf{F}_{M}\mathbf{r}))\oslash (\left|\mathbf{h}_{tf}\right|^{2}+\sigma^2),
\label{eq:eq21}
\end{equation}
where $\sigma^2$ is the noise variance, $\mathbf{h}_{tf} \triangleq \mathrm{diag}\left(F^H_M\mathbf{H_t}F_M\right)$ is the TF plane channel estimated by the proposed pilot, $\oslash$ denotes the Hadamard division.




\section{Numerical Results}
In this section we provide the performance evaluation from the perspective of both the sensing and communication. We assume the sensing and communication receivers are not necessarily to be co-located such that their channels are irrelevant. We assume a multi-target sensing, where the three targets correspond to three echo paths which follow a uniform distribution $U(0, \tilde{\nu}_{i})$ used to generate the Doppler shift, where $\tilde{\nu}_{i}$ is the maximum Doppler. For communications, we directly adopt the power delay profile (PDP) of 3GPP extended vehicular A (EVA) models for evaluating the symbol detection performance. We use the M-sequences generated by the different primitive polynomials as the component sequences of the 2D pilot. We investigate the performance of sensing via the Doppler estimation error and the performance of communication via the bit error probability (BER).
The detailed system setup is shown in table \ref{tab:tab0}.
\begin{table}
\begin{center}
\caption{System Parameters}
\label{tab:tab0}
\resizebox{\linewidth}{!}{
\begin{tabular}{|c||c||c|}
\hline
    Parameter & Communication & Sensing\\
\hline
    Carrier frequency & 6GHz & 6GHz\\ 
\hline
    SCS & 60e3 & 60e3\\ 
\hline
    [$N$,$M$] & [16,64] & up to [512,64]\\ 
\hline
    Power Scale & 1 & 0.2\\
\hline
    PDP &  3GPP EVA &  [0, 0, 0]\\ 
\hline
    Modulation & 16QAM & BPSK\\ 
\hline
    Codec & LDPC-0.33 & M-sequence\\ 
\hline
    Velocity & 30 & 500\\ 
\hline
\end{tabular}}
\end{center}
\end{table}

\begin{figure}
\centering
\includegraphics[width=0.7\textwidth]{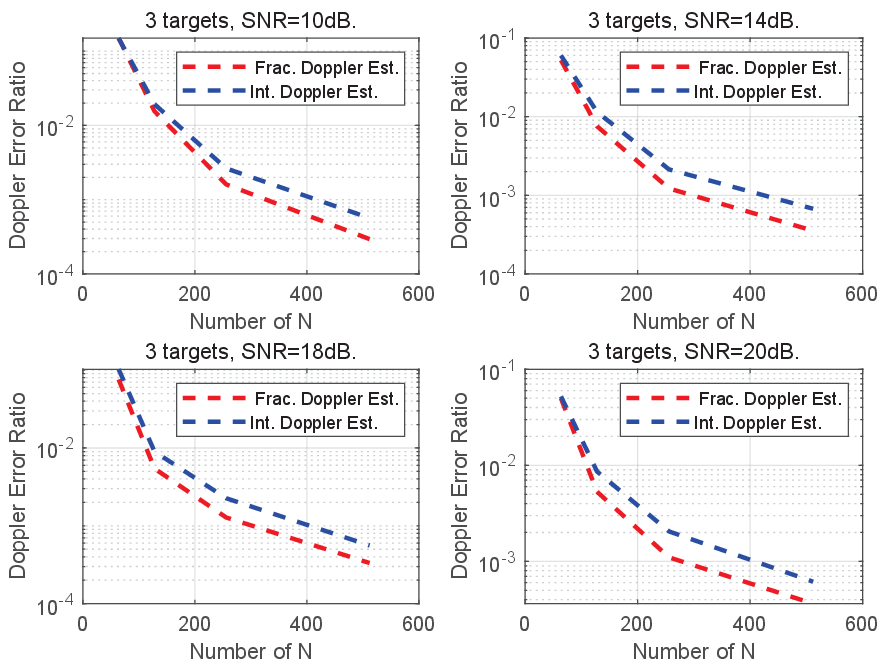}
  \caption[]{The comparison of sensing error in different pilot sizes.}
	\label{fig:fig3}
\end{figure}

Figure \ref{fig:fig3} shows the performance comparison of sensing detection performance between the correlation-based integer Doppler detection only and the former plus fractional Doppler refinement by magnitude information. Note that we define the Doppler error rate as 
$
\frac{1}{n}\sum^n_{i=1}{|\frac{\hat{D}_{i}-D_{i}}{D_{i}}|},
$
where $D_{i}$ and $\hat{D}_{i}$ are the Doppler and its estimation of target $i$ in an $n$-target detection. The three sub figures have SNRs of $10$dB, $14$dB, $18$dB and $20$dB, respectively, and we fix $M=64$ and increases $N$ from $64$ to $512$ to better demonstrates the impact of pilot size on Doppler accuracy.  
It is observed the estimation error decreases dramatically with increasing $N$, which coincides with our analysis in section \ref{sec:sec3}, i.e., increasing the size of the pilot can suppress the distortion of the sensing detection. 
By exploiting the magnitude information in refining the Doppler detection, the gap between the red and blue curves shows a promising gain of the proposed scheme against the integer Doppler detection. 

It is observed that the slope of the curves decrease rapidly when $N$ is small but slowly when $N$ is large. We can infer from this fact that the marginal gain of increasing the pilot size is bounded. However, the reason for the visibly worse performance of a small-size pilot is that the accumulated power of the correlation peak is too small to overwhelm the interfering data and noise, which can be negligible in the case the pilot is large. Furthermore, the stable performance with varying SNR shows the robustness of proposed scheme. 

\begin{figure}
\centering
\includegraphics[width=0.7\textwidth]{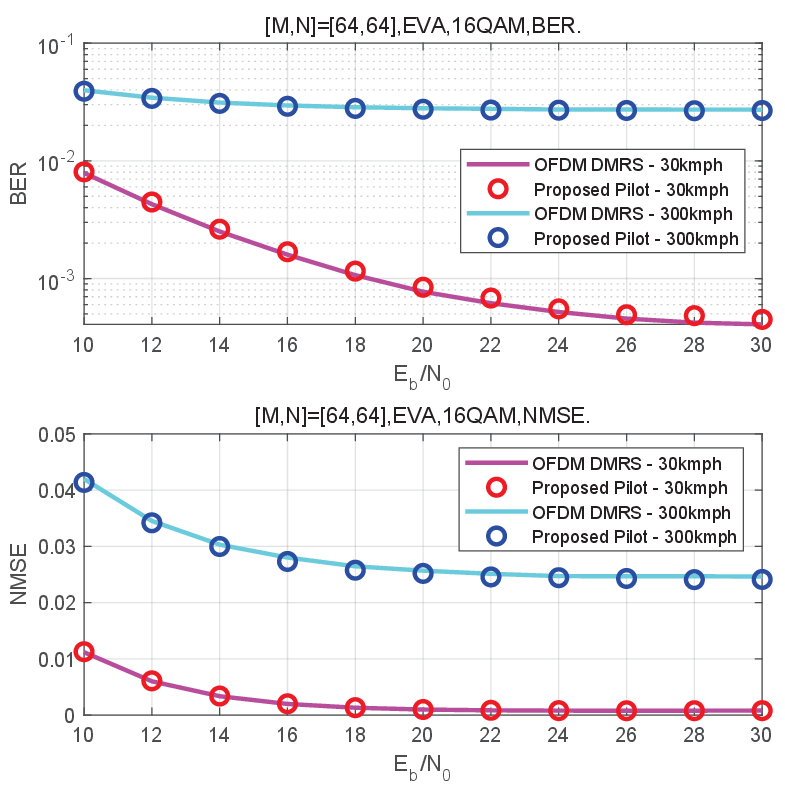}
  \caption[]{Communication performance: bit error probability and channel estimation error.}
	\label{fig:fig4}
\end{figure}

Figure \ref{fig:fig4} shows the comparison of the symbol detection performance of OFDM with conventional DMRS and the unified pilot. The power and resource allocation of the two is aligned for fairness. It is observed that the BER performance are almost identical except for high SNR region above $26$dB, where the BER performance of the unified pilot design is slightly worse. This is the consequent of the fact that the transformed pilot does not equally distribute the power among the samples due to the ISFFT, therefore some samples with low power gives less reliable estimation. 
It is also depicted in figure \ref{fig:fig4} that the channel estimation accuracy is maintained when replacing the conventional DMRS with the proposed one. It is observed that the performance in terms of normalized mean square error (NMSE) of conventional DMRS and unified design is not visibly different, as illustrated by the cyan solid line and magenta dotted curve with triangle. 

\section{Conclusion}
In this paper, we have presented and analyzed a unified pilot design for the ISAC systems. We have demonstrated that the proposed design is compatible with concurrent wireless communication systems and offers flexibility in adapting to different sensing requirements. Leveraging the twist-convolution property of the DD plane signal, we have provided a low-complexity sensing detection algorithm for integer delay and Doppler, and the fractional Doppler can be investigated by exploiting the magnitude information. We have shown that the change to the unified pilot design does not negatively impact the performance of the OFDM receiver. Our numerical results provide the evidence of the efficiency and robustness in distance and velocity estimation. Future work includes expanding the design to a multi-antenna system for angular estimation. 

\bibliographystyle{IEEEtran}
\bibliography{fmtc_TWC}
\end{document}